\documentclass[twocolumn,prb,showpacs,preprintnumbers,amsmath,amssymb,superscriptaddress]{revtex4}
\usepackage{graphicx}
\usepackage{color}
\usepackage{setspace}
\begin{document}

\title{Ordered phases in the Holstein-Hubbard model: Interplay of strong Coulomb interaction and electron-phonon coupling}
\author{Yuta Murakami}
\affiliation{Department of Physics, University of Tokyo, Hongo, Tokyo 113-0033, Japan}
\author{Philipp Werner}
\affiliation{Department of Physics, University of Fribourg, 1700 Fribourg, Switzerland}
\author{Naoto Tsuji}
\affiliation{Department of Physics, University of Tokyo, Hongo, Tokyo 113-0033, Japan}
\author{Hideo Aoki}
\affiliation{Department of Physics, University of Tokyo, Hongo, Tokyo 113-0033, Japan}
\date{\today}

\begin{abstract}
We study the Holstein-Hubbard model at half filling 
to explore ordered phases including superconductivity (SC), antiferromagnetism (AF), and charge order (CO) in situations where the 
electron-electron and electron-phonon interactions are strong ($\sim$ electronic bandwidth).
The model is solved 
in the dynamical mean-field approximation using a continuous-time quantum Monte Carlo impurity solver. 
We determine the superconducting transition temperature ($T_c$) and the SC order parameter 
and show that the phonon-induced retardation or the strong Coulomb interaction leads to a significant reduction and shift of the $T_c$ dome if one interprets the system as 
having an effective static interaction $U_\text{eff}$ given by the Hubbard $U$ reduced by the phonon-mediated attraction in the static limit.
This behavior is analyzed by comparison to an effective static model in the polaron representation with a reduced bandwidth.
We also determine the finite-temperature phase diagram including AF and CO.  In the moderate-coupling regime, there is a hysteretic region 
of AF and CO around $U_{\rm{eff}}=0$, while the two phases are separated by a paramagnetic metal in the weak-coupling regime and a paramagnetic insulator in the strong coupling regime.
\end{abstract}

\maketitle
\setstretch{0.9}
\section{\label{sec:level1}Introduction}
While the physics of correlated electron systems is an interesting 
and formidable problem in its own right, several classes of interesting materials exhibit an interplay of strong electron-electron Coulomb repulsion and strong electron-phonon coupling. For example, the electron-phonon coupling in high-$T_c$ cuprates is strong, as evidenced by the kinks observed in the angle-resolved photoemission spectrum.\cite{arpes} In alkali-doped fullerides an $s$-wave superconducting phase borders an antiferromagnetic phase in the temperature-pressure phase diagram\cite{c60_1,c60_2,c60_3} and the transition temperature of the superconducting state has recently been found to be dome-shaped. These features indicate that in these classes of materials, both the electron-electron 
Coulomb repulsion and the electron-phonon interaction are strong. Likewise, in aromatic superconductors such as picene (a recent addition to carbon-based materials), the electron-electron and electron-phonon interactions are reported to be strong,\cite{picene1,picene2,picene3} although the mechanism of superconductivity in the aromatic compounds is still 
totally unclear. Correlated electron systems often provide an interesting arena in which various phases compete with each other. In the presence of strong electron-electron and electron-phonon interactions, the problem should become even richer.

The Holstein-Hubbard (HH) model is a simple model which allows to describe and explore the interplay of electron-electron and electron-phonon interactions. The model incorporates a coupling between electrons and dispersionless (i.e., Einstein) phonons with energy $\omega_0$, in addition to the on-site Hubbard interaction $U$.  
There is a body of works which investigates the competition between the two interactions in this model. The study of the one-dimensional case based on the density matrix renormalization group (DMRG) technique or quantum Monte Carlo analysis has revealed some general features.\cite{1dimdmrg1,1dimdmrg3,sse1} 
However, since ordered phases with continuous symmetry breaking do not occur in $D=1$, it is difficult to elucidate the generic behavior of ordered states from these calculations, although we can indeed discuss quasi-ordered states in terms of the Tomonaga-Luttinger picture. In the opposite limit of infinite spatial dimensions, $D=\infty$, where the dynamical mean-field theory (DMFT) becomes exact,\cite{dmft2,dmft3,dmft4} ordered states with full symmetry breaking exist even at nonzero temperature. The effect of the competition between the two kinds of interactions on symmetric phases have been studied in Refs.~\onlinecite{HHdmft4,HHnrg,HHdmft6,HHnrg2}, and the corresponding phase diagram has been determined. 

As for the ordered states, their properties  have been investigated in several works,\cite{HHdmft1,HHdmft2,HHdmft3,Hol5,HHdmft7} but many issues remain unresolved. 
The ground state phase diagram around $U=\lambda$, where $\lambda$ is the static effective electron-electron interaction mediated by the phonons,
has been determined in Refs.~\onlinecite{HHdmft2,HHdmft3}, and deviations from the conventional theory of superconductivity have been discussed.\cite{Hol5,HHdmft7}  An important issue is 
the following: the electron-phonon coupled system is often regarded as having an 
effective interaction $U_\text{eff} \equiv U-\lambda$, but this is only strictly valid 
in the antiadiabatic limit for the phonon energy,  
$\omega_0\rightarrow\infty$, where the interaction in the HH model becomes 
non-retarded, and the real question is to what extent this approximation remains valid when we vary 
$U$ and/or $\omega_0$.   In other words, for a finite $\omega_0$ the phonon-mediated interaction is certainly retarded, and for small enough phonon frequency, the static model with $U_\text{eff}$ can be expected to fail.  Thus the nature of the superconducting state in the regime where $U$, $\lambda$, and $\omega_0$ are all comparable to the bandwidth $W$ poses an challenging problem, which is not only conceptually interesting, but may have relevance to real materials with strong electron-electron and electron-phonon interactions. 
The problem, however, has not been properly understood, since superconducting states in such a regime cannot be treated within conventional theories such as the Migdal theorem\cite{sc3,sc4} or the MacMillan equation.\cite{sc5} Another open issue is the finite-temperature phase diagram for ordered phases in the vicinity of $U=\lambda$.

With these questions in mind, we study in this paper ordered states in the half-filled Holstein-Hubbard model to clarify the effect of the coexistence of electron-electron and electron-phonon interactions on the $s$-wave superconducting state (SC), the antiferromagnetic (AF) state and the charge ordered state (CO). In our study we employ DMFT, with a continuous-time quantum Monte Carlo (CT-QMC) impurity solver, which is exact up to statistical errors, and can in principle access any parameter regime down to temperatures of about one percent of the 
electronic bandwidth. Furthermore, DMFT + CT-QMC can treat these three ordered states without bias, which enables us to systematically investigate their competition 
in the regime where $U$, $\lambda$, and $\omega_0$ are comparable to the bandwidth. Note that even though there have been several works discussing ordered states in HH model ,\cite{HHdmft1,HHdmft2,HHdmft3,HHdmft7} the present work is, to the best of our knowledge, the first attempt to directly treat these ordered phases at non-zero temperatures with DMFT+CT-QMC. For SC, we focus on the transition temperature and the superconducting order parameter. 
We show that the phonon-induced retardation (when 
the phonon energy is well below the antiadiabatic limit) 
or the Coulomb repulsion have the effect of significantly decreasing and 
shifting the $T_c$ dome against $U_\text{eff}$, and a similar shift occurs for the superconducting order parameter as well. 
In order to understand and interpret the observed behavior we use an effective static model  in a polaron representation with reduced bandwidth derived from a Lang-Firsov transformation, which has been introduced to investigate electron-phonon coupled systems in the strong-coupling or antiadiabatic regime.\cite{effmodel,effmodel2,effmodel3}  
We test the quantitative and qualitative reliability of the effective model by examining to what extent the model reproduces the transition temperature, superconducting order parameter and Green's function. As for AF and CO phases,  we determine the phase diagram at nonzero temperature, and show how these phases compete with each other at moderate to strong couplings.

The paper is organized as follows. In Sec.~\ref{Sec_model}, we introduce the Holstein-Hubbard model and explain how DMFT can deal with ordered phases of this model. We also derive the effective static model. In Sec.~\ref{Sec_results}, we discuss how the properties of the superconducting state depend on the parameters $U$, $\lambda$, and $\omega_0$ when these parameters are comparable to the bandwidth. We also show phase diagrams at nonzero temperatures around $U=\lambda$, and reveal how AF and CO compete with each other. Sec.~\ref{Sec_summary} gives a brief summary.

\section{Formalism}
\label{Sec_model}

\subsection{Model}

The Holstein-Hubbard (HH) model represents an electron system that is coupled to local (Einstein) phonons. The Hamiltonian is 
\begin{align}
 H=&-t\sum_{\langle i, j\rangle,\sigma}[c^{\dagger}_{j\sigma}c_{i\sigma}+H.c.]+\sum_i[Un_{i\uparrow}n_{i\downarrow}-\mu(n_{i\uparrow}+n_{i\downarrow})]\nonumber\\
 &+g\sum_i(b^{\dagger}_i+b_i)(n_{i\uparrow}+n_{i\downarrow}-1)+\omega_0\sum_ib^{\dagger}_ib_i,
  \label{eq:HHmodel}
\end{align}
where $i$, $j$ denote sites, $\sigma$ the spin, and the first sum is over nearest neighbors. $c^{\dagger}_{i,\sigma}$ denotes a creation operator of an electron, $b^{\dagger}_i$ a creation operator of a phonon, $t$ the hopping parameter, $U$ the on-site electron-electron interaction, $\mu$ the chemical potential, $g$ the coupling constant between electrons and phonons and $\omega_0$ the phonon frequency.  In this model, the phonon is envisaged as an optical mode with 
an approximately constant energy dispersion (Einstein model). 
Since the phonons are assumed to be noninteracting, one can integrate out the phonon part to derive the effective electron-electron interaction,
\begin{equation}
 U_{\mathrm{eff}}(\omega) = U-\frac{2g^2\omega_0}{\omega_0^2-\omega^2},
\end{equation}
in a path integral framework.
The effective interaction in the low-energy regime is thus 
\begin{equation}
U_{\mathrm{eff}}\equiv  U_{\mathrm{eff}}(\omega=0) \equiv U-\lambda, \quad
\lambda=2g^2/\omega_0.
\end{equation} 
If we take the anti-adiabatic limit of 
$\omega_0\rightarrow\infty$ with $\lambda$ and $U$ fixed, the HH model reduces to the Hubbard model with the interaction $U_{\mathrm{eff}}$. This low-energy effective interaction has been used as a measure of the characteristic effective net interaction in previous works.\cite{HHdmft2,HHdmft3,1dimdmrg1} 
In the present work we examine the validity of a static description in the parameter regime of interest, namely strong electron-electron and electron-phonon coupling, with phonon frequencies comparable to the bandwidth. We shall see that a proper static description involves the screened interaction, along with a reduced bandwidth.

In order to deal with superconducting (SC) states, we employ the Nambu formalism and define the local Green's function as 
\begin{equation}
 \begin{split}
 \hat{G}_{\mathrm{loc},i}(\tau)&\equiv-\langle T\Psi_{i}(\tau)\Psi^{\dagger}_{i}(0)\rangle\\
 &=
 \begin{bmatrix}
 G_{11,i}(\tau)& G_{12,i}(\tau)\\
  G_{21,i}(\tau)& G_{22,i}(\tau)
 \end{bmatrix},
 \end{split}
 \end{equation}
where $\Psi^{\dagger}_{i}\equiv(c_{i\uparrow}^{\dagger},c_{i\downarrow})$ are Nambu spinors.
We use  
\begin{equation}
\Phi=\langle c_{i\downarrow}c_{i\uparrow} \rangle_H=G_{12}(\tau=0_{+}),
\end{equation}
as the order parameter for the SC phase (assuming homogeneity), 
where $\langle\rangle_H$ denotes the equilibrium expectation value computed with the Hamiltonian $H$.

\subsection{Dynamical mean-field theory}

The dynamical mean-field theory (DMFT), which is exact in infinite spatial dimensions, \cite{dmft2,dmft3,dmft4} maps a lattice problem onto an effective impurity problem.  
When we take into account the superconducting state of the Holstein-Hubbard model, the Hamiltonian of the impurity problem is 
$H_{\rm{imp}}=H_{\mathrm{loc}}+H_{\mathrm{bath}}+H_{\mathrm{mix}}$, 
with the three terms
  \begin{align}
   H_{\mathrm{loc}}=&Un_{\uparrow}n_{\downarrow}
-\mu(n_{\uparrow}+n_{\downarrow})\nonumber\\
&+g(b^{\dagger}+b)(n_{\uparrow}+n_{\downarrow}-1)
   +\omega_0b^{\dagger}b,\label{H_imp1}\\
     H_{\mathrm{bath}}=&\sum_{\sigma,p}\epsilon_p c^{\dagger}_{p,\sigma}c_{p,\sigma}+\sum_{p}(\Delta_p c^{\dagger}_{p\uparrow}c^{\dagger}_{-p\downarrow}+{\rm H.c.})\label{H_imp2},\\
     H_{\mathrm{mix}}=&\sum_{\sigma,p}(V_p^{\sigma}d^{\dagger}_{\sigma}c_{p,\sigma}+{\rm H.c.}).
     \label{H_imp3}
  \end{align}
Here $d$ is the annihilation operator of the electron on the impurity, $n_{\sigma}$ the density of electrons with spin $\sigma$ on the impurity, $b$ the annihilation operator of a local phonon coupled to the impurity, and $c$ the annihilation operator of an electron in the bath, with the bath states labeled by the quantum number $p$. 
In a SC state the parameters $\Delta_p$ can be nonzero.
Thus the Hamiltonian describes an impurity that is coupled to local phonons in a superconducting bath. The bath ($H_{\mathrm{bath}}$) and mixing ($H_{\mathrm{mix}}$) terms are determined self-consistently in such a way that the impurity Green's function reproduces the local lattice Green's function of the HH model.  The only information on the lattice structure that enters a DMFT calculation is the density of states. Here we adopt the Bethe lattice, whose density of states is $\rho(\epsilon)=\frac{1}{\pi t}\sqrt{1-(\epsilon/(2t))^2}$. 
  We use the quarter of the bandwidth, $t$, as the unit of energy. 
 
Since we want to describe phases such as antiferromagnetic (AF) 
and charge-ordered (CO) ones with a broken $Z_2$ symmetry between 
sublattices, we introduce sub-lattice indices $\theta=A$, $B$ ($\bar{\theta}=B$, $A$) (here for the Bethe lattice, which is bipartite) and express the self-consistency equation in the form
\begin{equation}
 \begin{split}
 [\hat\Lambda_\theta]_{i,j}(\tau)&=-\langle TA_i(\tau) A_j^{\dagger}(0))\rangle_{H_{{\rm bath}}}\\
 &=[t^2\sigma_3\hat{G}_{\mathrm{loc}\bar{\theta}}(\tau)\sigma_3]_{i,j},
 \end{split}\label{eq:self_consis2}
 \end{equation} 
where $\hat \Lambda$ is the hybridization function of the impurity model. Here, $A_i=\sum_pV_p^{i}a_{p,i}$, $V^1_p=V_p^{\uparrow}$, $V_p^2=-V_p^{\downarrow}$, $(a_{p,1}^{\dagger}, a_{p,2}^{\dagger})\equiv(c_{p,\uparrow}^{\dagger},c_{-p,\downarrow})$ (all quantities for sublattice $\theta$), and $\sigma_3=\mathrm{diag}(1,-1)$ is a Pauli matrix. 
If we assume a homogenous system, as in the investigation of SC, then
 $\hat{\Lambda}_{\theta}(\tau)=\hat{\Lambda}(\tau)$
 is independent of the sublattice. 
%The hybridization function can be related to the impurity ``bath Green's function" by
%  \begin{equation}
%  \hat{G}_{0\theta}^{-1}(i\omega_n)=i\omega_n+\mu \sigma_3-
%\hat{\Lambda}(i\omega_n),
%\label{eq:self_consis1}
%  \end{equation}
%with $\omega_n=\pi T(2n+1)$.   
 
 The impurity problem, Eqs.~(\ref{H_imp1})-(\ref{H_imp3}) is solved with the continuous-time Quantum Monte Carlo impurity solver (hybridization expansion, i.e. we regard the mixing term (\ref{H_imp3}) as a perturbation term and perform a Monte Carlo sampling of the corresponding diagrammatic expansion),\cite{ctqmc1,ctqmc4} based on the method introduced in Ref.~\onlinecite{HHdmft4}. 
In this approach, a Lang-Firsov decoupling\cite{Lang} of the electrons and phonons and an analytical summation of all phonon contributions for each term in the expansion of the hybridization term enables an exact treatment of the quantum phonons.
In the present case, we extend this technique to impurity problems that couple to a superconducting bath. 
\footnote{For the attractive Hubbard model, the treatment of the superconducting phase within DMFT and the hybridization expansion CT-QMC approach has been discussed in Ref.~\onlinecite{ctqmc3}.}

\subsection{Effective static model}

Before presenting the DMFT results, let us first introduce an effective static model for the low-energy description of the HH model,\cite{effmodel,effmodel3} which is useful for discussing the properties of the SC phase. The first step in the derivation is to perform a Lang-Firsov (LF) canonical transformation for the HH model, $H_{\mathrm{LF}}=e^{S}He^{-S}$ with $S=\frac{g}{\omega_0}\sum_{i,\sigma}n_{i,\sigma}(b^{\dagger}_i-b_i)$. The explicit expression of $H_{\mathrm{LF}}$ is
\begin{equation}
\begin{split}
H_{\mathrm{LF}}&=-t\sum_{\langle i,j\rangle ,\sigma}[e^{\frac{g}{\omega_0}(b^{\dagger}_i-b_i)}e^{-\frac{g}{\omega_0}(b^{\dagger}_j-b_j)}c^{\dagger}_{i,\sigma}c_{j,\sigma}+H.c.]\\
&+U_{\mathrm{eff}}\sum_{i}n_{i,\uparrow}n_{i,\downarrow}+\mu_{\mathrm{eff}}\sum_i n_i+\omega_0\sum_{i}b^{\dagger}_ib_i,
\end{split}\label{eq:HLF}
\end{equation}
where $\mu_{\mathrm{eff}}=\mu-\frac{g^2}{\omega_0}$. 
Here, $c^\dagger$, after the LF transformation, has a meaning of creating a polaron, 
%in Eq.~(\ref{eq:HLF}), since if one goes back to the original Hamiltonian $H$,
%$c^\dagger$ is inverse transformed to $e^{-S}c^{\dagger}e^{S}=e^{-\frac{g}{\omega_0}(b^{\dagger}-b)}c^{\dagger}$, which creates an electron dressed by a phonon cloud in Eq.~(\ref{eq:HHmodel}).  
as is evident from the phonon factors. 
An effective low-energy model for the original fermions is obtained by assuming that the phonons are not much excited. In other words, the effective Hamiltonian for the fermion part is obtained as the projection onto the subspace of zero phonons, $H_{\mathrm{eff}}=\langle0|H_{\mathrm{LF}}|0\rangle$, where $|0\rangle$ is the vacuums of phonons. This description becomes exact in the limit where $\omega_0$ is large and the temperature is much lower than $\omega_0$. The Hamiltonian resulting from this projection is\cite{effmodel} 
  \begin{align}
  H_{\mathrm{eff}}=&-Z_Bt\sum_{\langle i,j\rangle,\sigma}[c^{\dagger}_{i,\sigma}c_{j,\sigma}+H.c. ]\nonumber \\
  &+U_{\mathrm{eff}}\sum_{i}n_{i,\uparrow}n_{i,\downarrow}+\mu_{\mathrm{eff}}\sum_i n_i, \nonumber \\
Z_B=&\exp(-g^2/\omega_0^2).
\label{eq:eff}
  \end{align}
This is nothing but the usual Hubbard model with a static interaction $U_{\mathrm{eff}}$ and a hopping parameter renormalized by $Z_B$. For this model we can readily derive physical quantities such as the transition temperature, the order parameter for the SC phase or the Green's functions from simulations of the  Hubbard model as follows.

Let us define $\Phi(T,U, U_{\mathrm{eff}}, Z_{\mathrm{B}})=\langle c_{\downarrow}c_{\uparrow}\rangle_H$ as the order parameter for the SC state. Within the effective model, the order parameter is expressed as
  \begin{equation}
  \begin{split}
    &\Phi(T,U, U_{\mathrm{eff}}, Z_{\mathrm{B}})=\langle e^{-2\frac{g}{\omega_0}(b^{\dagger}-b)}c_{\downarrow}c_{\uparrow}\rangle_{H_{\mathrm{LF}}}\approx\\& \langle 0|e^{-2\frac{g}{\omega_0}(b^{\dagger}-b)}|0\rangle \langle c_{\downarrow}c_{\uparrow}\rangle_{H_{\mathrm{eff}}}=Z_B^2\Phi_{0}(T/Z_{B},U_{\mathrm{eff}}/Z_{B})\\
    &\equiv \Phi_{\text{eff}}[Z_B,U_{\text{eff}}],\label{eq:deleff}
   \end{split}
  \end{equation}
where we have defined $\Phi_{0}(T,U)$ as the order parameter for the Hubbard model with hopping $t$ and interaction $U$ at temperature $T$.
  It follows that the transition temperature for the Holstein-Hubbard model ($T_c[U, U_{\mathrm{eff}}, Z_{B}]$) is related to 
that for the attractive Hubbard model ($T^0_{c}[U]$) by
  \begin{align}
  T_c[U, U_{\mathrm{eff}}, Z_{\mathrm{B}}]&\approx Z_\mathrm{B}\times T^0_{c}[U_{\mathrm{eff}}/Z_\mathrm{B}]\nonumber \\
  &\equiv T_{c,\text{eff}}[Z_B,U_{\text{eff}}]. \label{eq:Teff}
  \end{align}
  
In order to evaluate Green's functions, we again use the LF transformation,
  \begin{align}
 G_{\sigma}(\tau)=&-\langle T_{\tau} c_{\sigma}(\tau)c^{\dagger}_{\sigma}(0)\rangle_{H}\nonumber\\
 =&-\langle T_{\tau} e^{-\frac{g}{\omega_0}(b^{\dagger}(\tau)-b(\tau))}e^{\frac{g}{\omega_0}(b^{\dagger}(0)-b(0))} c_{\sigma}(\tau)c^{\dagger}_{\sigma}(0)
 \rangle_{H_{LF}}.\label{eq:greens_LF}
 \end{align}
Then we employ an approximation, which is to separate out the phonon factors from the expectation value with respect to $H_{LF}$
(in the situation where $\omega_0$ is large the phonon dynamics may be decoupled from the polaron dynamics). The fermionic part
is evaluated by the static approximation. This gives

  \begin{align}
 G_{\sigma}(\tau)
  \approx & -\langle  T_{\tau} c_{\sigma}(\tau)c^{\dagger}_{\sigma}(0)\rangle_{H_{\rm eff}}\nonumber\\
  &\times \langle T_{\tau} e^{-\frac{g}{\omega_0}(b^{\dagger}(\tau)-b(\tau))}e^{\frac{g}{\omega_0}(b^{\dagger}(0)-b(0))}\rangle_{H_{\rm ph}},\label{eq:greens_eff}
  \end{align}
  where $H_{\mathrm{ph}}=\omega_0\sum_ib^{\dagger}_ib_i$.
In other words, we treat the whole system as if its Hamiltonian is $H_{\mathrm{eff}}+H_{\mathrm{ph}}$.\cite{effmodel2}

  As for the anomalous part,
  \begin{align}
   G_{12}(\tau)\approx& -\langle T_{\tau} c_{\uparrow}(\tau)c_{\downarrow}(0)\rangle_{H_{\mathrm{eff}}}\nonumber\\
  &\times\langle T_{\tau} e^{-\frac{g}{\omega_0}(b^{\dagger}(\tau)-b(\tau))}e^{-\frac{g}{\omega_0}(b^{\dagger}(0)-b(0))}\rangle_{H_{\mathrm{ph}}}.\label{eq:offeff}
  \end{align}
  The phonon factor can be calculated analytically as 
  \begin{equation}  
  \begin{split}
  \langle &T_{\tau} e^{-s\frac{g}{\omega_0}(b^{\dagger}(\tau)-b(\tau))}e^{-s'\frac{g}{\omega_0}(b^{\dagger}(0)-b(0))}\rangle_{H_{\mathrm{ph}}}\\
 & =\exp\bigl\{-\frac{g^2/\omega_0^2}{e^{\beta\omega_0}-1}\bigl[(e^{\omega_0\beta}+1)+ss'(e^{\omega_0(\beta-\tau)}+e^{\omega_0\tau})\bigl]\bigl\},
 \end{split}\label{eq:phonon_cont}
  \end{equation}
  where $s,s'=\pm1$ and $0\le \tau\le \beta$.

\section{Results}
\label{Sec_results}

\subsection{Superconductivity}

Here, we investigate the SC phase at half-filling, in order to understand the effect of the retardation and the Coulomb repulsion on the SC state.
To focus on SC, we enforce the symmetries $G_{11}(\tau)=G_{11}(\beta-\tau)=G_{22}(\tau)=G_{22}(\beta-\tau)$,
which hold in the SC and normal states at half filling, but not in the AF and CO phases.
Strictly speaking, if we allow both CO and SC orders in the self-consistency loop, CO dominates over SC. 
Still, it should be meaningful to study the SC state at half-filling, since, 
if the system has a frustration (e.g. induced by a second-neighbor hopping on a bipartite lattice), CO and AF may be suppressed. The results of this subsection can be thought of as describing the properties of such frustrated systems.

%%%%%%%%%%%%%%%%%%%%%%%%%%%%%%%%%%%%%%%%%%%%%%%%%%%%%
\begin{figure}[h]
  \begin{center}
   \includegraphics[width=60mm]{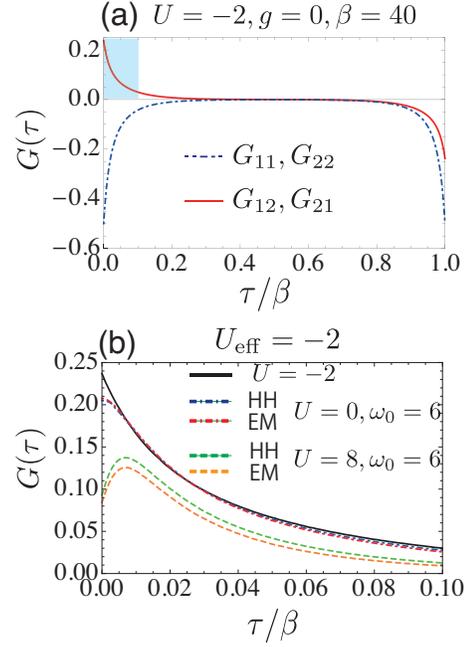}
  \end{center}
  \caption{(a) Typical behavior of Green's functions on the imaginary time 
axis, here for $U=-2$, $g=0$, and $\beta=40$.   
(b) A close-up (blue region in (a)) of 
the anomalous Green's function for various cases for fixed $U_{\mathrm{eff}}=
-2$ , where $g$ and $U$ are changed to keep $U_{\mathrm{eff}}=$ const with $\omega_0=6$ and $\beta=40$. `HH' means the Green's function is computed from the HH model, while `EM' means that it is obtained with the effective model. }
  \label{fig:greens_function}
\end{figure}
%%%%%%%%%%%%%%%%%%%%%%%%%%%%%%%%%%%%%%%%%%%%%%%%%%%%%

We first show that the Coulomb interaction induces a characteristic structure in the anomalous Green's function. Figure~\ref{fig:greens_function}(a) plots normal and anomalous Green's functions on the imaginary-time axis. While the diagonal Green's functions are negative and symmetric (at half-filling), the off-diagonal Green's functions are antisymmetric around $\tau/\beta=0.5$. 
In Fig.~\ref{fig:greens_function}(b) we show the short-time behavior of the anomalous Green's functions for different sets of parameter values: without retardation (Hubbard model with $U=-2$), with only a retarded attractive interaction ($U=0$ and $\omega_0>0$) and with both retardation and Coulomb repulsion. In all three cases, $U_{\mathrm{eff}}=-2$, $\omega_0=6$ and $T=0.025$. 
Without the retardation, the anomalous Green's function has its maximum at $\tau=0$. In the presence of a retarded attractive interaction, but without $U$, the position of the maximum remains at $\tau=0$, but the initial peak is rounded 
off. If we then switch on a $U>0$, the peak shifts to $\tau>0$, which indicates that when electrons form pairs, they tend to avoid the instantaneous repulsive interaction $U$ while exploiting the retarded attractive interaction.
 
One can explain the origin of this behavior with the effective model, Eq.(\ref{eq:offeff}). The corresponding results are also shown in Fig.\ref{fig:greens_function}(b).  
It turns out that the shift of the peak with $U$  in the anomalous Green's function is well reproduced by the effective model. This structure comes from the phonon part, Eq.~(\ref{eq:phonon_cont}), which increases with $\tau$ near $\tau=0$ and becomes steeper with $U$ for a fixed $U_{\mathrm{eff}}$.

 %%%%%%%%%%%%%%%%%%%%%%%%%%%%%%%%%%%%%%%%%%%%%%%%%%%%%
  \begin{figure}[h]  
     \begin{center}
   \includegraphics[width=70mm]{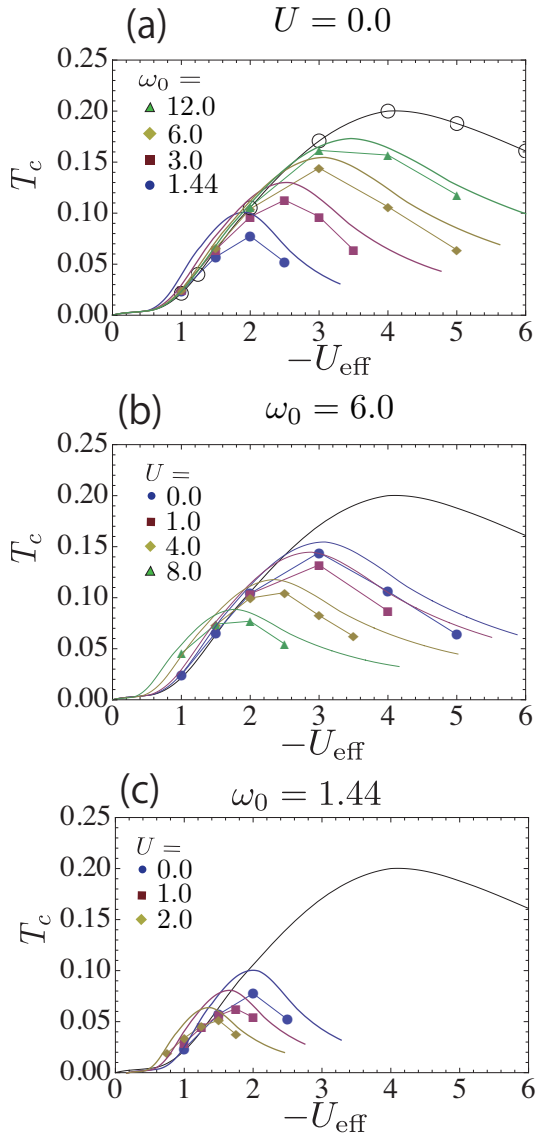}
  \end{center}
  \caption{ $T_C$ against $-U_{\mathrm{eff}}$ for various sets of parameter values. (a) shows the dependence on $\omega_0$ for $U=0$. The lines connecting symbols are guides for the eye. The unfilled circle show the results for the attractive Hubbard model, and the black curve indicates $T_c$ of it.  (b), (c) show the phase diagram when $U$ is switched on with a fixed value of $\omega_0=6$ (b), or $\omega_0=1.44$ (c). 
The colored curves show the results from the effective model. }

  \label{fig:phase_sc}
  \end{figure}
  %%%%%%%%%%%%%%%%%%%%%%%%%%%%%%%%%%%%%%%%%%%%%%%%%%%%%
Next, we clarify the effect of the two different interactions on the phase diagram. In Fig.~\ref{fig:phase_sc} we plot $T_c$ as a function of $-U_{\mathrm{eff}}$. Panel (a) illustrates the effect of the retardation (controlled by $\omega_0$) on the SC phase for the case $U=0$. As $\omega_0$ decreases, the position of the peak in $T_c$ shifts to the weak-$|U_{\mathrm{eff}}|$ regime, while the height of the peak decreases. We also note that the $T_c$ in the weak-coupling regime still agrees well with that of the attractive Hubbard model with interaction $U_{\mathrm{eff}}$.  Let us compare this results with the behavior of CO in the Holstein model. \cite{Hol1,Hol2} The shift of the $T_c$  dome with $\omega_0$ is also observed in CO, while the height of the $T_c$ peak does not show significant  change. Further, the transition temperature increases in the weak-coupling regime when $U_\text{eff}$ is fixed and $\omega_0$ decreases in CO.\cite{Hol1,Hol2}
The difference between SC and CO in the weak-coupling region can be explained as follows. The important interaction for SC is the interaction between electrons with opposite spins. For CO, however, the phonon mediated interaction between electrons with the same spin is also relevant, as can be understood from a mean-field analysis in the adiabatic limit (see Appendix). 
When the phonon frequency is reduced, the lattice distortion takes so much time that an electron starts to feel the attraction from another electron having the same spin through this distortion. Due to this additional attraction, $T_c$ increases in CO when $\omega_0$ decrease, while SC does not take this advantage.

In Fig.~\ref{fig:phase_sc}(b) and (c), the effect of the Coulomb repulsion is illustrated. As $U$ increases, the position of the peak shifts to the weak-coupling regime and the height of the peak decreases. It turns out that the $T_c$ in  the weak-coupling region is not necessarily well reproduced by the attractive Hubbard model with interaction strength $U_{\mathrm{eff}}$ (see e.~g. $U=8$ in panel(b) or $U=2$ in panel (c)).

Let us now examine the above properties in terms of the low-energy effective static model in the polaron representation, which is a Hubbard model with interaction $U_{\mathrm{eff}}$ {\it and} a renormalized hopping parameter (reduced by the factor $Z_B$, see Eq.~(\ref{eq:eff})).  
The resulting transition temperature qualitatively well describes the dependence of the transition temperature on $\omega_0$, $U$, and $\lambda$, see colored curves in Fig.~\ref{fig:phase_sc}.  
The effective model somewhat overestimates the transition temperature. An increase in $U$ or a decrease in $\omega_0$ with $U_{\mathrm{eff}}$ fixed leads to an increase of $Z_B=\exp(-\frac{\lambda}{2\omega_0})$, since $\lambda=U-U_\text{eff}$. 
Therefore, the band renormalization for the polaron enhances the effect of interactions between polarons, which may be characterized by the ratio between $U_{\mathrm{eff}}$ and the bare polaron band width $Z_BW$, where $W=4t$ is the band width of bare electrons. As a result, the peak of the transition temperature shifts to smaller $|U_{\mathrm{eff}}|$. Note that the effective model also shows that with lager $U$, the deviations from the attractive Hubbard model (black lines in Fig.~\ref{fig:phase_sc}) increase in the weak-coupling regime, see for example the colored lines for  $U=8$ in Fig.~\ref{fig:phase_sc}(b) or $U=2$ in Fig.~\ref{fig:phase_sc}(c). 
This is related to the fact that the shape of $T_c$ in the attractive Hubbard model is convex 
in the weak-coupling (BCS) regime. Therefore, the enhancement of the correlation due to the renormalization of the hopping parameter by $Z_B$ can lead to deviations from the attractive Hubbard model if we do not rescale the $U$-axis. 
Here one may wonder why, when $\omega_0$ is changed with $U=0$ and $U_\text{eff}$ fixed,
 the deviation from the attractive Hubbard model is not so apparent than when $U$ is changed with $\omega_0$ and $U_\text{eff}$ fixed.
 This is because we need smaller $\omega_0$ to realize a given value of $Z_B$ when $U=0$, while the reliability of the effective model is degraded for smaller $\omega_0$.

%%%%%%%%%%%%%%%%%%%%%%%%%%%%%%%%%%%%%%%%%%%%%%%%%%%%% 
   \begin{figure}[h]  
     \begin{center}
   \includegraphics[width=70mm]{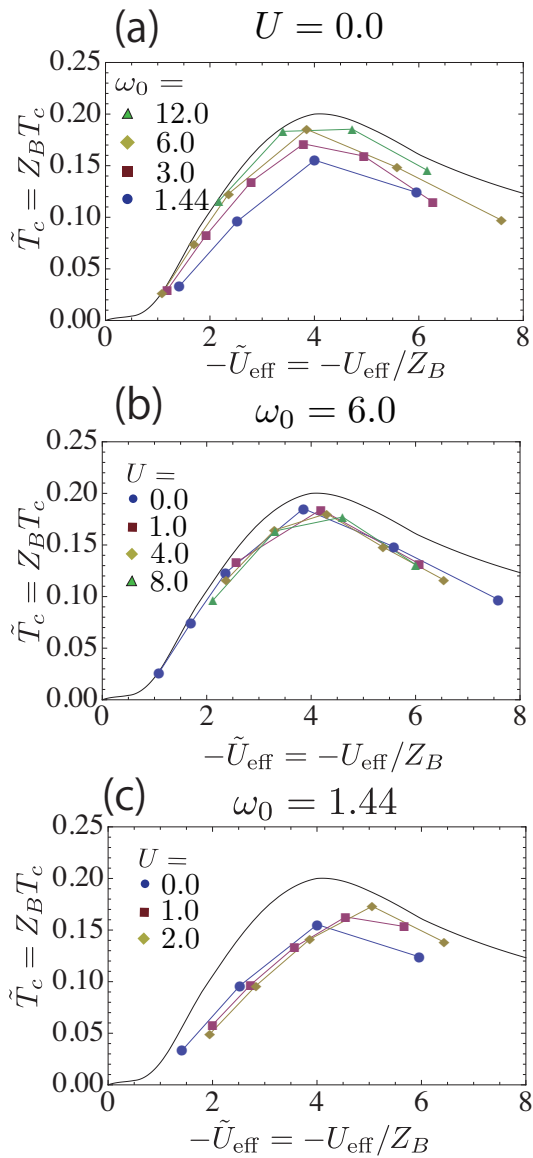}
  \end{center}
  \caption{ The phase diagram with rescaled parameters: $\tilde{U}_{\text{eff}}\equiv U_{\mathrm{eff}}/Z_B$ and $\tilde{T}\equiv Z_BT$. (a) shows the result for various 
values of $\omega_0$ at $U=0$. (b) (c) are the result when $U$ is switched on with a fixed $\omega_0=6$ (b), or $\omega_0=1.44$ (c). A black curve in each panel shows $T_c$ in the attractive Hubbard model. }
  \label{fig:phase_sc2}
  \end{figure}
  %%%%%%%%%%%%%%%%%%%%%%%%%%%%%%%%%%%%%%%%%%%%%%%%%%%%% 
  In order to estimate in which region the effective model agrees with the Holstein-Hubbard model, 
we plot the phase diagram in terms of the rescaled 
$\tilde{U}_{\text{eff}}\equiv U_{\mathrm{eff}}/Z_B$ {\it and} 
rescaled $\tilde{T}\equiv Z_BT$ in Fig.~\ref{fig:phase_sc2}. 
If the effective model reproduces the result of the Holstein-Hubbard model, the phase diagram in the space of $\tilde{T}$ and $\tilde{U}_{\text{eff}}$ should coincide
with that of the attractive Hubbard model.  
Our numerical results cover the range $6\agt\tilde U_{\text{eff}}\agt2$. As expected, the reliability of the effective model becomes better as $\omega_0$ increases (Fig.~\ref{fig:phase_sc2} (a)). For $U=0$, the relative deviation ($\delta T_c\equiv |T_c-T_{c,\text{eff}}|/T_c$, where $T_{c,\text{eff}}|$ is defined in Eq.~\ref{eq:Teff}. )
  is smallest at intermediate coupling ($\tilde{U}_{\text{eff}}\sim4$) for each $\omega_0$. 
The deviation decreases from $\delta T_c \leq 0.25$ for $\omega_0=4$ 
to $\delta T_c \leq 0.1$  for $\omega_0=12$. 
   The dependence of $\delta T_c$ on $U_{\mathrm{eff}}$ is relatively small at $\omega_0\geq 4$. On the other hand, if $\omega_0\leq 2$, the reliability of the effective model deteriorates with increasing $U_{\mathrm{eff}}$, as shown in Fig.~\ref{fig:phase_sc2}(a). As for the effect of $U$, we find that $\delta T_c$ slightly, but systematically increases with increasing $U$, at least in the weak-coupling regime (panels (b) and (c)), and  at $\omega_0=4$, $\delta T_c \leq 0.25$ up to $U=4$.  
The effective model is quantitatively accurate up to larger values of $U$ for larger $\omega_0$. 
  %%%%%%%%%%%%%%%%%%%%%%%%%%%%%%%%%%%%%%%%%%%%%%%%%%%%%
         \begin{figure}[h]  
     \begin{center}
   \includegraphics[width=70mm]{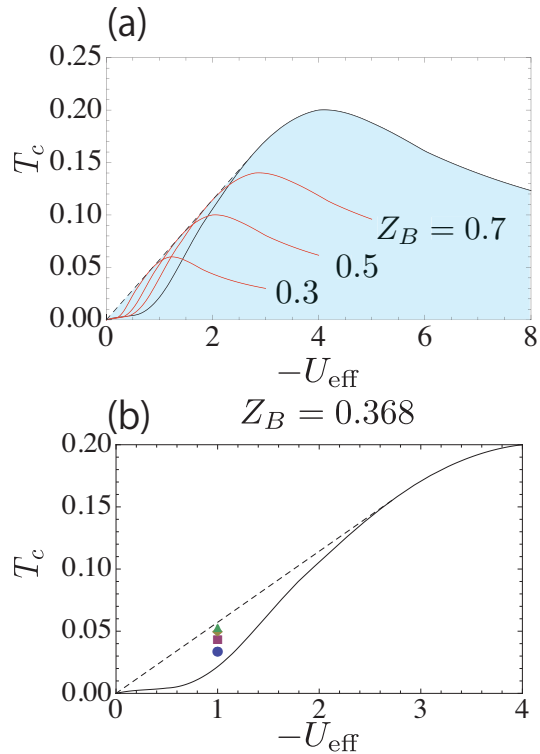}
  \end{center}
  \caption{(a) Possible area (shaded) for the transition temperature in the Holstein-Hubbard model, which is an envelope of the $T_c$ curves (red) for various 
values of $Z_B$. 
(b) $T_c$ vs $-U_{\mathrm{eff}}$ for various values of 
$\omega_0$, plotted here for $U_{\mathrm{eff}}=-1, Z_B=U_{\mathrm{eff}}/U_0$. }
  \label{fig:z_fix}
  \end{figure}
%%%%%%%%%%%%%%%%%%%%%%%%%%%%%%%%%%%%%%%%%%%%%%%%%%%%%

 The above analysis enables us to discuss the region in the  $T$ versus $-U_{\mathrm{eff}}$ space where a superconducting phase of the Holstein-Hubbard model 
can exists. The shaded area in Fig.~\ref{fig:z_fix}(a) shows this region as predicted by the effective model. The boundary of this area defines an envelope for the various $T_c$ curves, which 
rises from the origin linearly and touches the $T_c$ curve of the attractive model before the peak. This is because a set of $T_c$ curves for various values of 
$Z_B$ form a homologous series of phase boundaries of the attractive Hubbard model, Eq.~(\ref{eq:Teff}). In the weak-coupling regime of the Hubbard model, the phase boundary is convex, while the boundary is concave in the intermediate regime, where the BCS-BEC crossover occurs. 
The envelope indeed becomes a tangent to
the original curve (black solid line in Fig. \ref{fig:z_fix}) at 
$U_{\mathrm{eff}}\simeq -2.72\equiv U_0$.  The boundary of the blue area is obtained by fixing $U_{\mathrm{eff}}$, $Z_B=U_{\mathrm{eff}}/U_0$
and taking the limit $\omega_0\rightarrow\infty$, see Fig.~\ref{fig:z_fix}. 
The effective model always overestimates the transition temperature in the 
parameter region studied here, so that we expect that the superconducting phase of the Holstein-Hubbard model is contained within the blue area. Here, we note that when we take the limit $\omega_0\rightarrow\infty$ with $U_{\mathrm{eff}}$, $Z_B$ fixed, $U_{\mathrm{eff}}(\omega)\rightarrow U_{\mathrm{eff}}$ for every finite $\omega$, but because $\lambda\rightarrow \infty$, this does not mean that the Holstein-Hubbard model becomes the attractive Hubbard model without bandwidth reduction. 
     
 %%%%%%%%%%%%%%%%%%%%%%%%%%%%%%%%%%%%%%%%%%%%%%%%%%%%%       
       \begin{figure}[h]  
     \begin{center}
   \includegraphics[width=70mm]{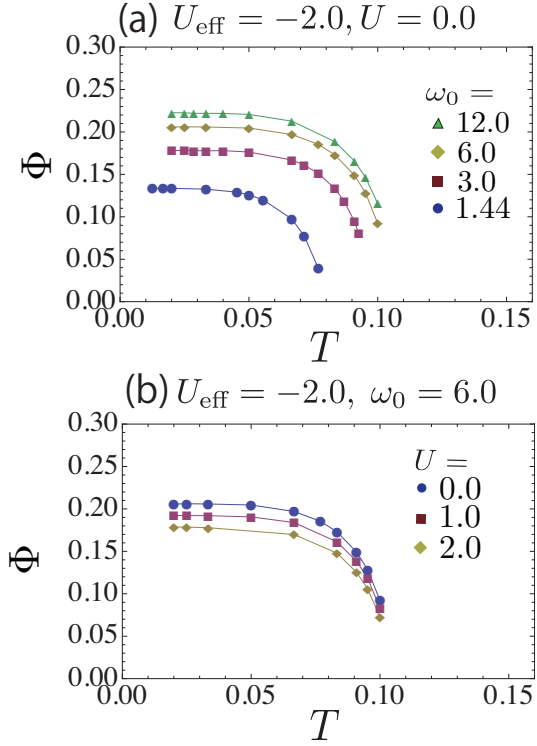}
  \end{center}
  \caption{Temperature dependence of the superconducting order parameter ($\Phi(T)$) 
for $U_{\mathrm{eff}}=-2, U=0$ with various values of $\omega_0$ (a), 
or for $U_{\mathrm{eff}}=-2,\omega_0=6$ with various  values of $U$ (b). }
  \label{fig:scdens}
  \end{figure}
%%%%%%%%%%%%%%%%%%%%%%%%%%%%%%%%%%%%%%%%%%%%%%%%%%%%% 

Next we discuss the properties of the superconducting state itself. Here we focus on the 
temperature dependence of the superconducting order parameter, $\Phi(T)$.  
In Fig.~\ref{fig:scdens}(a) we fix $U=0$, $U_{\mathrm{eff}}=-2$ and change the value of $\omega_0$. In Fig.~\ref{fig:scdens}(b) we fix $\omega_0=6$, $U_{\mathrm{eff}}=-2$ and change $U$. We find that $\Phi$ monotonically increases below $T_c$, and saturates as temperature is decreased. 
As can be seen in panels (a) and (b), the retardation and the Coulomb repulsion $U$ both act to decrease $\Phi(T)$. 

%%%%%%%%%%%%%%%%%%%%%%%%%%%%%%%%%%%%%%%%%%%%%%%%%%%%% 
   \begin{figure}[h]  
     \begin{center}
   \includegraphics[width=70mm]{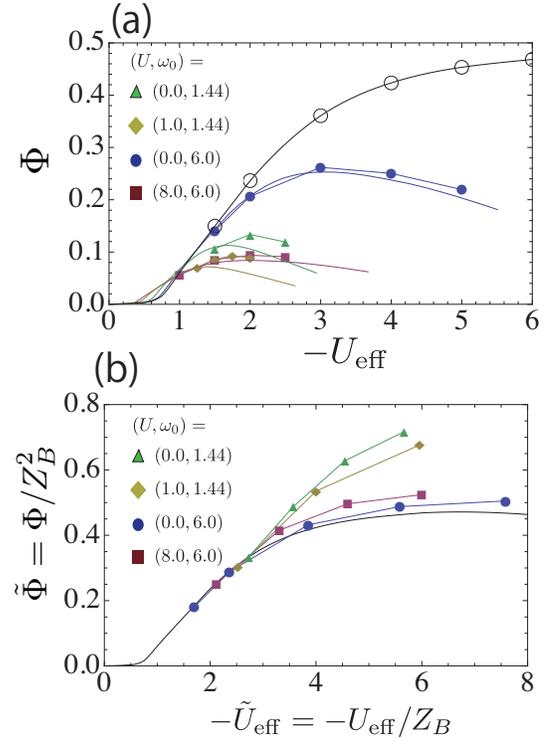}
  \end{center}
  \caption{The superconducting order parameter at $T=0$ ($\Phi(T \rightarrow 0)$). Panel (a) plots the result against $-U_{\mathrm{eff}}$ for various sets of parameter values. The unfilled circles are  $\Phi(T \rightarrow 0)$  for the attractive Hubbard model, and the black curve show the interoperation of them. Panel (b) plots the result against rescaled quantities: $\tilde{U}_{\mathrm{eff}}\equiv U_{\mathrm{eff}}/Z_B$ and $\tilde{\Phi}\equiv \Phi/Z_B^2$.}
  \label{fig:delta}
  \end{figure}
  %%%%%%%%%%%%%%%%%%%%%%%%%%%%%%%%%%%%%%%%%%%%%%%%%%%%%  
 In order to investigate the effect of the two types of interactions on the superconducting order parameter more systematically, we focus on the value of $\Phi$ in the limit of $T\rightarrow 0$ (Fig.~\ref{fig:delta}). In panel (a),  $\Phi(T \rightarrow 0)$ is plotted as a function of $-U_{\mathrm{eff}}$. In the attractive Hubbard model, $\Phi(T \rightarrow 0)$ saturates at 0.5
in the strong coupling limit.\cite{ctqmc3} On the other hand, for finite $\omega_0$, we find that it has a peak as a function of $-U_{\mathrm{eff}}$. Furthermore, the peak shifts to  smaller $|U_{\mathrm{eff}}|$ as the retardation increases ($\omega_0$ decreases) or the Coulomb interaction $U$ increases. 
We also note that, in the region investigated ($-U_{\mathrm{eff}}\agt 1.5$), $\Phi(T \rightarrow 0)$ decreases as $\omega_0$ decreases or $U$ increases, as illustrated in Fig.~\ref{fig:scdens}. It turns out that this behavior is qualitatively well described by the effective model, see the color lines in Fig.~\ref{fig:delta}(a).

We can also explain the origin of the peak structure in $\Phi(T \rightarrow 0)$ as follows. Within the effective model, what saturates at large $U_{\rm{eff}}$
 for $\omega_0\neq0$ is the density of pairs of polarons, which can be expressed as $\langle c_{\downarrow} c_{\uparrow}\rangle_{\mathrm{LF}}$ after the Lang-Firsov transformation, while 
$\Phi$, the order parameter defined for electrons, has some correction coming from the phonon dressing. This correction becomes large as the electron-phonon coupling becomes large, see Eq.~(\ref{eq:deleff}) ($Z_B$ decreases as $\lambda$ increases). 
Related to the discussion of Fig.~\ref{fig:scdens}, 
we also have to note that the effective model, in the weak-coupling region, predicts that there is some area where $\Phi(T \rightarrow 0)$ increases as $\omega_0$ decreases or $U$ increases. However, within our approach (CT-QMC based on the hybridization expansion), this region is difficult to access, since it is in the weak-coupling regime and at very low temperature.

  In order to assess the potential of the effective model to reproduce the superconducting order parameter, we rescale the axis of Fig.~\ref{fig:delta}(a) 
into $\tilde{U}_{\text{eff}}\equiv U_{\mathrm{eff}}/Z_B$ and $\tilde{\Phi}\equiv \Phi/Z_B^2$, and show the result in Fig.~\ref{fig:delta}(b).  Again we focus on the range $2\lesssim -\tilde{U}_{\mathrm{eff}}\lesssim6$. It turns out that, with large enough $|\tilde{U}_{\text{eff}}|$, 
the rescaled curve underestimates $\Phi(T \rightarrow 0)$, while 
for smaller $|\tilde{U}_{\text{eff}}|$ ($\lesssim 2.5$), the effective model becomes better. 
  A larger $U$ leads to a larger underestimation in the strong $|\tilde{U}_{\text{eff}}|$ regime. Quantitatively , $\delta\Phi \equiv |(\Phi(T \rightarrow 0)-\Phi_\text{eff}(T \rightarrow 0))|/\Phi(T \rightarrow 0) \leq 0.2$ for $\omega_0\agt 4$ at $U=0$, where $\Phi_\text{eff}$ is defined in Eq.~(\ref{eq:deleff}).  
  As for the effect of $U$, we find $\delta\Phi \leq 0.2$ up to $U =6$ at $\omega_0= 4$. The effective model is quantitatively accurate up to larger $U$ for larger $\omega_0$.
  The reliability of the effective model for $\Phi$ is slightly better than in the case of the transition temperature.

   Now let us move on to discuss the energy gap in the spectral function and its relation with the transition temperature. We express the self-energy, which is independent of momentum in DMFT, as
   \begin{equation}
   \hat{\Sigma}(i\omega_n)
   =
 \begin{bmatrix}
 \Sigma(i\omega_n)& S(i\omega_n)\\
  S(i\omega_n)& -\Sigma^*(i\omega_n)
 \end{bmatrix},
   \end{equation}
where $\Sigma$ is the normal self-energy while $S$ is the anomalous one.   
Then the lattice Green's function in momentum space is
   \begin{equation}
   \begin{split}
    &\hat{G}({\bf k}, i\omega_n)=
  \begin{bmatrix}
G_{11}({\bf k}, i\omega_n)& G_{12}({\bf k}, i\omega_n)\\
  G_{21}({\bf k}, i\omega_n)& G_{22}({\bf k}, i\omega_n)
 \end{bmatrix}\\
&=[|G^{0}({\bf k}, i\omega_n)^{-1}-\Sigma(i\omega_n)|^2+|S(i\omega_n)|^2]^{-1}\times
 \\&
   \begin{bmatrix}
 G^{0}(-{\bf k}, -i\omega_n)^{-1}-\Sigma(-i\omega_n)& -S(i\omega_n)\\
  -S(i\omega_n)& -G^{0}({\bf k},i\omega_n)^{-1}+\Sigma(i\omega_n)
 \end{bmatrix},
 \end{split} \label{eq:full_g}
 \end{equation}
 where $G^{0}({\bf k},i\omega_n)=i\omega_n-(\epsilon_{\bf{k}}-\mu)$ is the bare lattice Green's function and $\epsilon_{\bf{k}}$ the dispersion relation for the bare electrons.
 With Dyson's equation and Eq.~(\ref{eq:self_consis2}), one finds for the Bethe lattice that
 \begin{equation}
   \hat{\Sigma}(i\omega_n)=i\omega_n\sigma_0+\mu\sigma_3-t^2\sigma_3 \hat{G}_\text{loc}(i\omega_n) \sigma_3-\hat{G}_\text{loc}^{-1}(i\omega_n),
 \end{equation}
where $\sigma_0$ is the identity matrix.
 As long as the Fermi liquid picture is valid, we can express the self-energies for small $|\omega_n|$ as
 \begin{align}
  &\Sigma(i\omega_n)=\Sigma^{(0)}+i\omega_n \Sigma^{(1)}+O((i\omega_n)^2),\\
  &S(i\omega_n)=S^{(0)}+O((i\omega_n)^2), \label{eq:qp_approx}
 \end{align}
 where $\Sigma^{(0)}$, $\Sigma^{(1)}$, $S^{(0)}$ are real, and $Z=(1-\Sigma^{(1)})$ denotes the Matsubara-axis mass-renormalization factor. 
 Neglecting the $O((i\omega_n)^2)$ term in Eq.~(\ref{eq:qp_approx}), inserting the expansions into Eq.~(\ref{eq:full_g}) and making an analytic continuation, one finds that the gap in the excitation spectrum is 
 \begin{equation}
 \Delta\equiv S^{(0)}/Z.
 \end{equation}
 This provides a rough estimate of the spectral gap. In the BCS theory, the ratio between the energy gap and the transition temperature ($2\Delta(T \rightarrow 0)/T_c$) is $3.528$, so that any deviation from this value is a measure for the deviation from the BCS theory.

\begin{figure}[h]  
     \begin{center}
   \includegraphics[width=70mm]{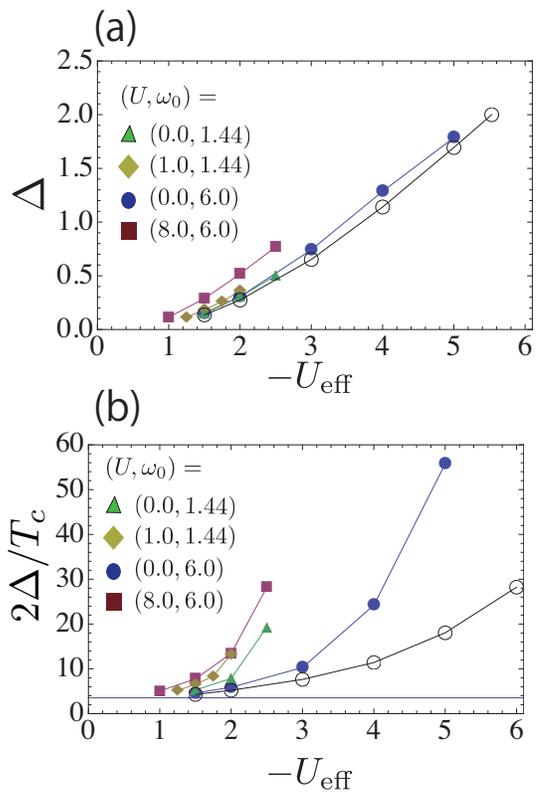}
  \end{center}
  \caption{(a) The energy-gap parameter $\Delta (=\Delta(T \rightarrow 0)$) 
against $-U_{\mathrm{eff}}$ for various sets of parameters.  (b) The ratio of the energy-gap parameter to the transition temperature 
($2\Delta(T \rightarrow 0)/T_c$) against $-U_{\mathrm{eff}}$. The horizontal 
blue line indicates the BCS value $2\Delta/T_c=3.528$. The unfilled circles are the results  for the attractive Hubbard model for each panel.}
  \label{fig:gap}
  \end{figure}

We plot these quantities in Fig.~\ref{fig:gap}. Here we use the approximations $\Sigma^{(1)}\approx \rm{Im}\Sigma{(\omega_{n=0})}/\omega_{n=0}$, and $S^{(0)}\approx (9S{(\omega_{n=0})}-S{(\omega_{n=1})})/8$. The gap parameter for $T\rightarrow0$, $\Delta(T \rightarrow 0)$, is shown in Fig.~\ref{fig:gap}(a). It turns out that it monotonically increases with $|U_{\rm{eff}}|$. When we decrease $\omega_0$ for fixed $U=0$, $\Delta(T \rightarrow 0)$ does not necessarily change monotonically with $\omega_0$ for each value of $U_{\mathrm{eff}}$,
but the dependence on $\omega_0$ is small. When we increase $U$ with fixed $\omega_0$, it increases for each $-U_{\rm{eff}}$.

Fig.\ref{fig:gap}(b) plots 
$2\Delta(T \rightarrow 0)/T_c$. Again it monotonically increases with $|U_{\mathrm{eff}}|$.  We find that, while 
$2\Delta(T \rightarrow 0)/T_c$ approaches  the BCS value $3.528$ for $-U_{\mathrm{eff}} 
\rightarrow 0$ in all the cases, there is a steep rise at a 
value of $|U_{\mathrm{eff}}|$ that depends on $\omega_0$ and $U$.  
The deviation increases when $\omega_0$ is decreased or $U$ is increased.
   
   %%%%%%%%%%%%%%%%%%%%%%%%%%%%%%%%%%%%%%%%%%%%%%%%%%%%% 
  %%%%%%%%%%%%%%%%%%%%%%%%%%%%%%%%%%%%%%%%%%%%%%%%%%%%%
  %%%%%%%%%%%%%%%%

  Finally let us discuss the possible relevance of these results for organic superconductors, such as alkali-doped fullerides and aromatic superconductors.\cite{c60_3, picene1,picene2}  
 In these molecular solids, the characteristic frequency of intra-molecular phonons 
is reported to be not larger than but comparable to the electronic bandwidth or the inverse of the density of states at the Fermi level. 
What we have found here is that the effective static model is qualitatively reliable even for $\omega_{0}\simeq t=W/4$. 
Therefore, we expect that the static effective model in the polaron picture is useful to analyze the qualitative behavior, such as the dependence on pressure. We caution however that in a realistic study of organic superconductors, one has to consider multiple orbitals and different types of electron-electron and electron-phonon couplings (Hund's couplings, intra-orbital phonon couplings and/or Jahn-Teller interactions).

   \subsection{Antiferromagnetism and charge order}
   
To understand the competition of ordered phases in the presence of two kinds of interactions, we now investigate the Holstein-Hubbard model at non-zero temperatures around $U_{\mathrm{eff}}=0$, without any constraint (i.e., allowing SC, commensurate CO and commensurate AF). In the following, the order parameter of the CO phase ($\Phi_{\rm{CO}}$) is defined as $\Phi_{\rm{CO}} = [(n_{A,\uparrow}+n_{A,\downarrow}) - 
(n_{B,\uparrow}+n_{B,\downarrow})]/4$, 
where $n_{A,\sigma}$ ($n_{B,\sigma}$) represents the density of electrons on the $A$ ($B$) sublattice with spin $\sigma$. For the AF phase the order parameter is defined as $\Phi_{\rm{AF}}=(n_{\uparrow}-n_{\downarrow})/2$, where $n_{\uparrow} (n_{\downarrow})$ is the density of up-spin (down-spin) electrons.
To obtain the data points, we use the hybridization functions for $U_{\mathrm{eff}}=U-\lambda$ as an input for the next step 
$U_{\mathrm{eff}}=U-\lambda-\delta\lambda$, where $\delta\lambda$ denotes 
a small increment in $\lambda$. 
    %%%%%%%%%%%%%%%%%%%%%%%%%%%%%%%%%%%%%%%%%%%%%%%%%%%%% 
     \begin{figure}[h]  
     \begin{center}
   \includegraphics[width=60mm]{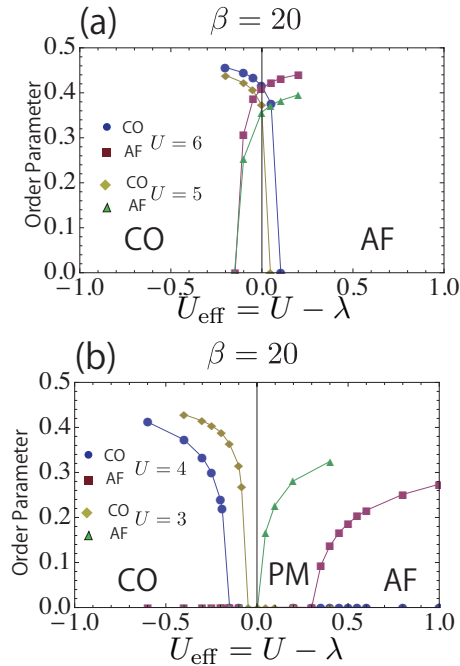}
  \end{center}
  \caption{Order parameters against $U_{\rm eff} = U-\lambda$ around $U\sim\lambda$ for a fixed $\omega_0=0.6$ at half-filling. In panel (a), the result for $U=6$, $5$ is shown, while in panel (b) the result for $U=4,3$, 
both at $\beta=20$. For smaller $U$, the charge ordered state (CO) and the antiferromagnetic phase (AF) are separated by a paramagnetic metallic phase (PM).}
  \label{fig:order_half}
  \end{figure}
%%%%%%%%%%%%%%%%%%%%%%%%%%%%%%%%%%%%%%%%%%%%%%%%%%%%%
%%%%%%%%%%%%%%%%%%%%%%%%%%%%%%%%%%%%%%%%%%%%%%%%%%%%%  
       \begin{figure}[h]  
     \begin{center}
   \includegraphics[width=70mm]{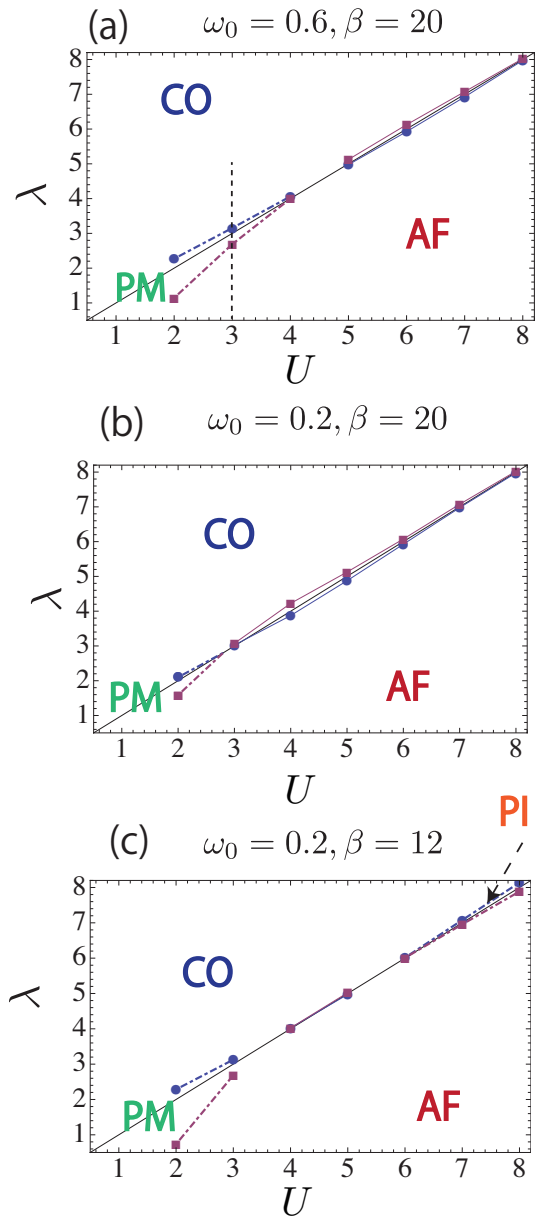}
  \end{center}
  \caption{Finite temperature phase diagram in the $U$-$\lambda$ plane.  
On top of charge ordered state (CO), antiferromagnetic phase (AF), CO, AF and 
paramagnetic metallic phase (PM), PI stands for a paramagnetic insulating state. Colored dotted lines represent the boundary between a paramagnetic phase and ordered phases, which is continuous. Solid lines show the boundary of the region where the stable solution for CO or AF exists. For each type of lines, the red line shows the boundary of AF and blue line is for CO. }
  \label{fig:phase_af_co}
  \end{figure}
  %%%%%%%%%%%%%%%%%%%%%%%%%%%%%%%%%%%%%%%%%%%%%%%%%%%%%
    
Figure~\ref{fig:order_half} shows the behavior of the order parameters around $U_{\mathrm{eff}}=0$, which has been obtained by varying $\lambda$ to change $U_{\mathrm{eff}}$ for each value of $U$. Here we fix $\omega_0=0.6$ and $\beta=20$. There is no SC phase, and AF and CO compete with each other. 
If the interaction $U$ is strong enough ($U=5$, $6$, see Fig.~\ref{fig:order_half}(a)), 
the phase transition is of first order 
i.e., the order parameters show a hysteresis around the phase boundary. 
In other words,  there is a region around $U_{\mathrm{eff}}=0$, in which both an AF and a CO solution of the DMFT equations exists. In order to determine the stable solution, one would have to compare the free energies.
On the other hand, if the interaction $U$ is smaller ($U=3$, $4$, see Fig.~\ref{fig:order_half}(b)), a paramagnetic metallic phase (PM) appears between the CO and AF phase. The transition to PM is second order, since the order parameter continuously goes to $0$ as one approaches the boundary Fig. \ref{fig:order_half}.

\begin{figure}[h]  
     \begin{center}
   \includegraphics[width=70mm]{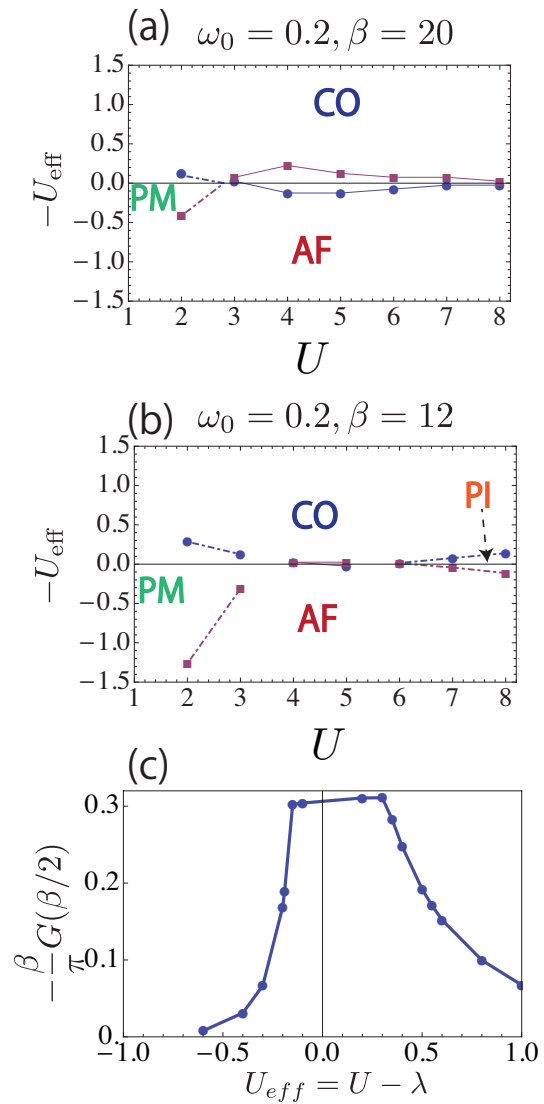}
  \end{center}
  \caption{(a)(b)Finite temperature phase diagram in the $U$-$-U_{\rm{eff}}$ plane.  
 Colored dotted lines represent the boundary between a paramagnetic phase and ordered phases, which is continuous. Solid lines show the boundary of the region where the stable solution for CO or AF exists. For each type of lines, the red line shows the boundary of AF and blue line is for CO.
 (c)$-\frac{\beta}{\pi}G(\tau=\beta/2)$ against $U_{\mathrm{eff}}$ 
with a fixed $U=3$ (vertical dotted line in Fig.\ref{fig:phase_af_co}(a)) with $\mu=1.5,\;\omega_0=0.6$.   }
  \label{fig:af_co_detail}
  \end{figure}

We can summarize the results by plotting the phase diagrams in the $U$-$\lambda$ plane for various conditions in Fig.~\ref{fig:phase_af_co} and by plotting the phase diagrams in the $U$-$-U_{\rm{eff}}$ in Fig.~\ref{fig:af_co_detail}(a)(b). In the weak coupling regime, as pointed out above, there is a paramagnetic metallic phase (PM) 
around $U_{\mathrm{eff}}=0$. It turns out that the area of the PM phase is wider on the $U>\lambda$ side than on the opposite side. This can be explained by a mean-field theory in the adiabatic limit, which has been used to explain the first-order transition in the strong coupling regime at $T=0$ in Ref.~\onlinecite{HHdmft3}. This approximation shows that 
the interaction that appears in the gap equation 
is $U-2\lambda$ for CO and $U$ for AF, as elaborated in the Appendix, and this explains the different extents of the two regions. The paramagnetic metallic phase between AF and CO is also expected in the anti-adiabatic limit, where the HH model becomes the Hubbard model with interaction $U_{\mathrm{eff}}$. Ref.~\onlinecite{HHdmft3} pointed out that the nature of this limit shows up in the weak-coupling regime as a continuous transition between AF an CO. In that sense, the present result is consistent with Ref.~\onlinecite{HHdmft3}.
However, we note that the Hubbard model cannot explain the difference in the extent of the PM region between $U>\lambda$ and $U<\lambda$, since the Hubbard model solution should be symmetric around the line $U=\lambda$. 
In the intermediate-coupling regime, the transition between AF and CO is of first-order and takes place within the hysteretic region, see Fig.~\ref{fig:af_co_detail}(a)(b) for more details.    
This hysteretic region of two solutions (AF and CO) is located near $U\sim\lambda$ even at non-zero temperatures, and is shown as the region surrounded by red and blue solid lines in Fig.~\ref{fig:phase_af_co} and in Fig.~\ref{fig:af_co_detail}(a)(b).  This result agrees with the results at $T=0$.\cite{HHdmft2,HHdmft3}
We also find that, in the strong-coupling regime, the hysteretic region becomes narrower with larger $U$ (or $\lambda$) (Fig.~\ref{fig:phase_af_co} (a)(b), Fig.~\ref{fig:af_co_detail}(a)). In the large-$U$ regime, the CO and AF solutions are separated by a paramagnetic insulating phase (Fig.~\ref{fig:phase_af_co}(c), Fig.~\ref{fig:af_co_detail}(b)). 
When we compare the different panels, we find that  the region of paramagnetic states is suppressed as the temperature decreases (compare Figs.~\ref{fig:phase_af_co}(b) and (c)) or as $\omega_0$ decreases (compare Figs.~\ref{fig:phase_af_co}(a) and (b)). The latter may be because the cancellation of the instantaneous repulsive interaction and the retarded interaction is more direct than in the case of smaller $\omega_0$.

%%%%%%%%%%%%%%%%%%%%%%%%%%%%%%%%%%%%%%%%%%%%%%%%%%%%%  
  %%%%%%%%%%%%%%%%%%%%%%%%%%%%%%%%%%%%%%%%%%%%%%%%%%%%%

Finally, we show the evidence for the metallic nature of the PM phase in the small-$U$ regime. We use the relation between the Green's function on the imaginary axis and the spectral function ($A(\omega)=-\frac{1}{\pi}\mathrm{Im} G_{\mathrm{loc}}(\omega)$),
\begin{equation}
G(\tau=\beta/2)=-\int  d\omega \frac{1}{2\cosh(\beta\omega/2)}A(\omega).
\end{equation}
If the temperature is low enough, $1/\cosh(\beta\omega/2)$ has a strong peak at $\omega=0$. Then the value of $-\frac{\beta}{\pi}G(\tau=\beta/2)$ gives a good estimate of the value of  $A(\omega=0)$. 
The result is shown in Fig.~\ref{fig:af_co_detail}(c) for $U=3$, $\mu=1.5$, and indicates that there is indeed a significant density of states at the Fermi level in the PM phase. Note that we can only observe a small value of $-\frac{\beta}{\pi}G(\tau=\beta/2)$ in the paramagnetic state in the strong coupling regime, so that this phase must be regarded as a paramagnetic insulating phase.

 \section{Conclusion}
 \label{Sec_summary}
 We have systematically investigated the effect of the electron-electron interaction and the electron-phonon coupling on the ordered states of the half-filled Holstein-Hubbard model, using DMFT and CT-QMC. In the study of the superconducting state, we have found that the interplay of the Coulomb interaction and the retarded attractive interaction leads to a nontrivial structure in the anomalous Green's functions, and we have shown that the maximum transition temperature deceases as a result of the electron-phonon interaction and shifts to the small $U_{\mathrm{eff}}$ regime. The superconducting order parameter shows a similar behavior. We have explained these observations with an effective static model derived from a Lang-Firsov decoupling and a projection onto the zero-boson subspace, and we have investigated the accuracy and reliability of the effective model. We have also  discussed the region where a SC state can be realized in the HH model in the $T_c$-versus-$U_{\mathrm{eff}}$ phase diagram. Finally, we have investigated the HH model at $T>0$ around $U_{\mathrm{eff}}=0$, allowing for SC, AF and CO phases. A paramagnetic phase (PM) appears between CO and AF in the weak-coupling region and a paramagnetic insulating phase (PI) for strong enough coupling, while in the intermediate-coupling regime, the transition between CO and AF is direct and discontinuous and a hysteresis region of AF and CO is located around $U_{\mathrm{eff}}=0$ at non-zero temperatures.  
 
\acknowledgements
The authors would like to thank T. Kariyado, T. Oka, M. Tezuka and A. Koga for constructive advice and discussion.
 The simulations in this study have been performed using some of the ALPS libraries (Ref.~\onlinecite{ALPS}).

\appendix
\section{Static mean-field treatment for AF and CO}
We briefly discuss a static mean-field theory for a heuristic understanding of the behavior of CO and AF. We treat both interactions in Eq.~(\ref{eq:HHmodel}) with a mean-field approximation by introducing the average of the lattice displacement $\langle b^{\dagger}_i+b_i\rangle/\sqrt{2\omega_0}$ and the averaged density $\langle n_{i,\sigma}\rangle$. The mean-field Hamiltonian is decomposed as
\begin{equation}
H_{\mathrm{MF}}\equiv H^{e}_{{\rm MF}}+ H^{ph}_{{\rm MF}},
\end{equation}
where 
\begin{equation}
\begin{split}
& H^{e}_{\mathrm{MF}}=-t\sum_{\langle i, j\rangle ,\sigma}[c^{\dagger}_{j,\sigma}c_{i,\sigma}+H.c.]\\
 &+\sum_i[U(\langle n_{i\uparrow}\rangle n_{i\downarrow}+\langle n_{i\downarrow}\rangle n_{i\uparrow})-\mu n_{i}+g\langle b^{\dagger}_i+b_i\rangle n_{i}],
\end{split}
\end{equation}
and 
\begin{equation}
 H^{ph}_{\mathrm{MF}}=g\sum_i(b^{\dagger}_i+b_i)(\langle n_{i}\rangle-1)+\omega_0\sum_ib^{\dagger}_ib_i.
\end{equation}
From $ H^{ph}_{\mathrm{MF}}$, we find $\langle b_i \rangle=\langle b^{\dagger}_i \rangle =-\frac{g}{\omega_0}(\langle n_i \rangle -1)$. Then we obtain
\begin{equation}
\begin{split}
 H^{e}_{\mathrm{MF}}&=-t\sum_{\langle i, j \rangle,\sigma} [c^{\dagger}_{j,\sigma}c_{i,\sigma}+H.c.]-\sum_i(\mu-\lambda)n_i\\
 &+\sum_{i,\sigma}n_{i\sigma}(U\langle n_{i\bar{\sigma}}\rangle -\lambda \langle n_i \rangle ).
 \end{split}
\end{equation}
This Hamiltonian shows that the effective attractive interaction, $-n_{i\sigma}\lambda\langle n_i \rangle$, comes from electrons of both spins, which is different from the case of the Coulomb interaction. In the following, let us focus on half-filling and consider the solution of CO and AF.

For CO, $\langle n_{A\uparrow}\rangle=\langle n_{A\downarrow}\rangle\neq \langle n_{B\uparrow}\rangle=\langle n_{B\downarrow}\rangle$. We define the order parameter as $\Phi_{\rm{CO}}=(\langle n_{A\uparrow}\rangle-\langle n_{B\uparrow}\rangle)/2$.   On the other hand, for AF, $\langle n_{A\uparrow}\rangle=\langle n_{B\downarrow}\rangle\neq \langle n_{A\downarrow}\rangle=\langle n_{B\uparrow}\rangle$. We define the order parameter as $\Phi_{\rm{AF}}=(\langle n_{A\uparrow}\rangle-\langle n_{A\downarrow}\rangle)/2$. 
Then we obtain a self-consistent equation,
\begin{equation}
 1=V\int d \xi \rho(\xi)\frac{\tanh(\beta E(\xi,\Phi, V)/2)}{2E(\xi,\Phi, V)}, \label{eq:gap_eq}
 \end{equation}
 where $\rho(\xi)$ is the density of states for bare electrons and $E(\xi,\Phi,V)=\sqrt{V^2\Phi^2+\xi^2}$. For CO we put $\Phi=\Phi_{\rm{CO}}, V=|U-2\lambda|$ and for AF,  $\Phi=\Phi_{\rm{AF}}, V=|U|$.
 
Note that if we consider CO and SC in the attractive Hubbard model within the static mean-field approximation, we put $V=|U|=|U_{\rm{eff}}|$ in Eq.~(\ref{eq:gap_eq}). On the other hand, when we consider the Holstein model, we use $V=2|\lambda|=2|U_{\rm{eff}}|$, which corresponds to the effective interaction for both spin up and down. This explains why the transition temperature for CO is enhanced in the small $U_{\rm{eff}}$ regime as $\omega_0$ decreases, which was pointed out in Refs.~\onlinecite{Hol1,Hol2}.

We also note that the static mean-field analysis can explain the reason why the PM region is larger on the $U>\lambda$ side than on the $U<\lambda$ side. This comes from the different dependence of AF and CO on $U$ and $\lambda$.  
At a given temperature $T$, let $V_0>0$ satisfy
\begin{equation}
 1=V_0\int d \xi \rho(\xi)\frac{\tanh(\beta E(\xi,0, V_0)/2)}{2E(\xi,0, V_0)}. 
\end{equation}
The mean-field analysis then dictates that the boundary of CO and PM is located at $\lambda=(V_0+U)/2$, while the boundary of AF and PM is at $U=V_0$. The two boundaries cross at $U=\lambda=V_0$, and a first-order transition occurs for $U,\lambda>V_0$.

\end{document}